\documentclass[pra,
reprint,showpacs,amsmath,amssymb,nofootinbib]{revtex4-1}

\usepackage{graphicx}

\usepackage{amsmath,amsthm,amsfonts,amssymb,bm}
\usepackage{xcolor} 
\usepackage[colorlinks=false,urlcolor=purple, linkcolor=purple]{hyperref}
\usepackage{subfigure}
\usepackage{centernot}
\usepackage{mathtools}
\usepackage{stmaryrd}

\def\({\left(}
\def\){\right)}

\def\eq#1{Eq.~(\ref{eq:#1})}
\def\Eq#1{Equation~(\ref{eq:#1})}
\def\eqs#1#2{Eqs.~(\ref{eq:#1}-\ref{eq:#2})} 
\def\eqlist#1#2{Eqs.~(\ref{eq:#1}-\ref{eq:#2})}

\def\fig#1{Fig.~\ref{fig:#1}}
\def\figs#1#2{Figs.~\ref{fig:#1} \& \ref{fig:#2}}
\def\Fig#1{Figure~\ref{fig:#1}}

\newcommand{\disp}{D}

\DeclareMathOperator{\sinc}{sinc}

\begin{document}

\title{Engineering large-scale entanglement in the quantum optical frequency comb: influence of the quasiphasematching bandwidth and of dispersion} 
\author{Pei Wang}
\email{pw4cq@virginia.edu}
\author{Wenjiang Fan}
\author{Olivier Pfister}
\email{opfister@virginia.edu}
\affiliation{Department of Physics, University of Virginia, Charlottesville, Virginia 22903, USA}


\begin{abstract}
One of the most scalable methods for continuous-variable quantum computing is to encode quantum information in the quantum optical frequency comb generated by an optical parametric oscillator (OPO). The scalability is limited by the quasiphasematching bandwidth and dispersion of the OPO. We study both factors in this article. The results show that 3200 qumodes are probably entangled in our recent demonstration of a record 60-qumode entanglement.
\end{abstract}


\maketitle 

\section{Introduction}
Quantum computing~\cite{Ladd2010} can be implemented, among other options, with quantum optics~\cite{Lloyd1999,Bartlett2002,Menicucci2006,Menicucci2014ft} and in particular using, in lieu of qubits, the continuous variables defined by individual quantum oscillator modes of the electromagnetic field~\cite{Braunstein2005a,Weedbrook2012}. These {\em qumodes} are characterized by their quadrature amplitude operators which are the analogues of the position and momentum of a harmonic oscillator $\hat q=(\hat a+\hat a^\dag)/\sqrt2$, $\hat p=(\hat a-\hat a^{\dag})/(i\sqrt2)$, where $\hat a$ is the photon annihilation operator of a particular mode. Operators { $\hat q$ and $\hat p$} have a continuous spectrum. An interesting particular implementation uses the quantum optical frequency comb (QOFC) of an optical parametric oscillator (OPO)~\cite{Pfister2004,Menicucci2008}, whose important feature is its ``top-down'' scalability potential. This has been demonstrated experimentally by entangling a world-record 60  modes of the QOFC of a single OPO~\cite{Pysher2011,Chen2014}. (Note that 10,000 modes were also entangled sequentially in a different work, but were only accessible 2 modes at a time~\cite{Yokoyama2013}.) In the QOFC experiments, the OPO modes are spaced by a FSR of 0.95 GHz and the OPO's nonlinear crystal provides entanglement by way of two-mode squeezing. The number of entangled modes is therefore given by the ratio of the phasematching bandwidth of the OPO crystal  and the FSR. When the frequency span is large, due to dispersion, the modes generated by two-mode squeezing will not be enhanced by the OPO cavity eventually. So the total number of modes are limited by both quasiphasematching bandwidth and dispersion.

In this work, we studied, theoretically and experimentally, quasiphasematched parametric downconversion (PDC) with a monochromatic pump in periodically poled KTiOPO$_{4}$ (PPKTP) pumped at 532 nm. The experimental study used sum-frequency generation. The OPO dispersion is studied by theoretical simulation. We show that the quasiphasematching is very broad and flat, with a 4.74 THz theoretical full width, at 99\% of the maximum efficiency, and a measured full width of 3.18 THz, which yields 6700 entangled OPO modes of both orthogonal polarizations after compensating the OPO dispersion.

\section{Quasiphasematching Bandwidth}

\subsection{Simple model}

\begin{figure}[hb!]
\centerline{\includegraphics[width=\columnwidth]{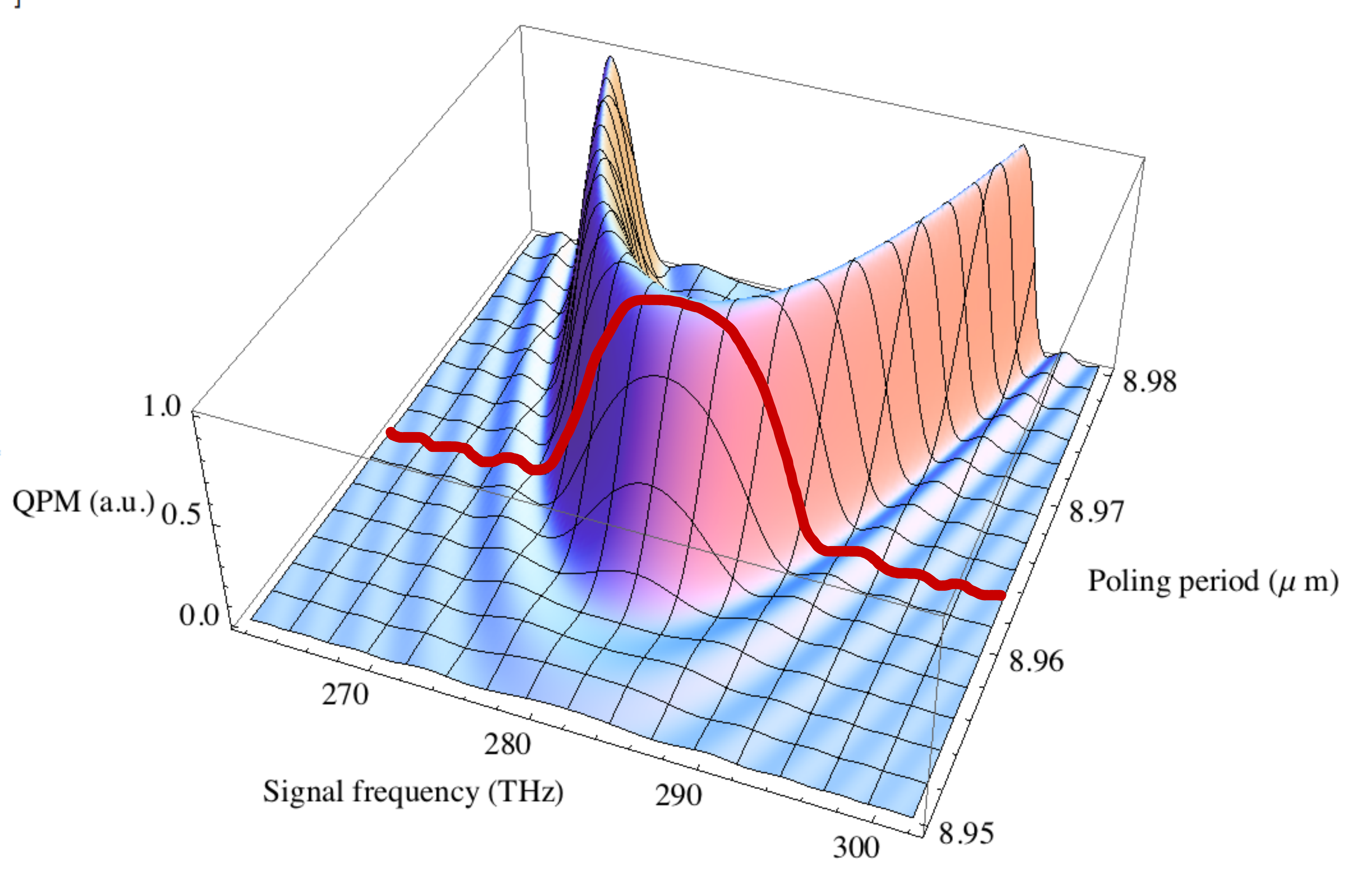}}
\caption{Quasiphasematching function, versus signal frequency $\omega_{1}$ and poling period, for a 1 cm-long PPKTP crystal. The red line outlines the considered QPM FWHM bandwidth, which reaches 4.74 THz at 99\% of the maximum and decreases significantly in the separated branches.}
\label{fig:th}
\end{figure}

We modeled  quasiphasematched (QPM)~\cite{Armstrong1962,Fejer1992,Myers1995}  $zzz$ single sum-frequency generation (SFG) to the exact, fixed pump frequency in PPKTP. This is the reverse process to PDC in an OPO pumped by a monochromatic field, as used in our entanglement experiments, but its phasematching bandwidth is identical and it lends itself more easily to experimental characterization. We followed the standard plane-wave treatment in the slowly varying amplitude approximation. All classical fields are of the form $E_j=A_je^{-i(k_jx-\omega_{j}t)}$, $j\in[1,5]$. Two monochromatic, linearly polarized input fields $E_{1,2}$ at frequencies $\omega_{1,2}$ travel in the crystal along its $x$ axis and give rise to SFG field $E_3$ at $\omega_{3}=\omega_{1}+\omega_{2}$, as well as SHG fields $E_{4,5}$ at $2\omega_{1}$ and $2\omega_{2}$. The coupled wave equations are
\begin{align}
{da_1\over dx}&=-ig_1a_1^*a_4e^{-i\Delta k_1x}-ig_3a_2^*a_3e^{-i\Delta k_3x} \label{eq:1}\\
{da_2\over dx}&=-ig_2a_2^*a_5e^{-i\Delta k_2x}-ig_3a_1^*a_3e^{-i\Delta k_3x} \\
{da_3\over dx}&=-ig_3a_1a_2e^{i\Delta k_3x} \\
{da_4\over dx}&=-i{g_1\over 2}a_1^2e^{i\Delta k_3x} \\
{da_5\over dx}&=-i{g_2\over 2}a_2^2e^{i\Delta k_3x} \label{eq:5}
\end{align}
where $a_j=A_j/\sqrt{2\eta_j\hbar\omega_j}$, $\eta_j=\eta_0/n_j$ is the impedance of the medium, $g_{1,2}^2=4\hbar\omega_{1,2}^3\eta_{1,2}^2\eta_{4,5}d_{33}^2$, $g_3^2=2\hbar\omega_1\omega_2\omega_3\eta_1\eta_2\eta_3d_{33}^2$, and $d_{33}$ is the second-order nonlinear coefficient. The SHG and SFG phase mismatches are, respectively, 
\begin{align}
&\Delta k_{1,2}=n(2\omega_{1,2},T){2\omega_{1,2}\over c}-2n(\omega_{1,2},T){\omega_{1,2}\over c}-{2\pi\over\Lambda},\\
&\Delta k_3=n(\omega_3,T){\omega_{3}\over c}-n(\omega_1,T){\omega_{1}\over c}-n(\omega_2,T){\omega_{2}\over c}-{2\pi\over\Lambda},\label{eq:sfg}
\end{align}
where $\Lambda$ is the poling period. In the limit of undepleted pumps $a_{1,2}$, \eqlist{1}{5} can be solved simply to yield the total intensity
\begin{align}\label{eq:int}
I = & g_3^2L^2|a_1a_2|^2\sinc^2\Bigl({\Delta k_3L\over2}\Bigr)+{g_{1}^2\over 4}L^2|a_{1}|^4\sinc^2\Bigl({\Delta k_1L\over2}\Bigr) \nonumber\\
& +{g_{2}^2\over 4}L^2|a_{2}|^4\sinc^2\Bigl({\Delta k_2L\over2}\Bigr).
\end{align}
We first examine the SFG phasemismatch of \eqs{int}{sfg} at room temperature as a function of the poling period $\Lambda$ and of SFG input --- or OPO ``signal'' --- frequencies $\omega_{1}$ and $\omega_{2}=\omega_{3}-\omega_{1}$, $\omega_{3}$ being fixed to the pump frequency (563 THz, 532 nm) of the OPO. This can be done by using { temperature-dependent} Sellmeier equations~\cite{Fradkin1999,Emanueli2003}. The result, plotted in \fig{th}, 
displays a remarkably broad phasematching frequency bandwidth around frequency degeneracy ($\omega_{1}=\omega_{2}=\omega_{3}/2$). Therefore, we can expect a 532 nm pumped, $zzz$ PPKTP OPO to achieve mode entanglement over a 4.74 THz full width at 99\% maximum, which would yield 10,000 entangled cavity modes for an OPO of  0.95 GHz FSR.

Analogous calculations for longer wavelengths, such as  775/1550 nm, indicate that the QPM bandwidth at 99\% of the maximum should be broader by a factor of 3. 

\subsection{Experiment and results}

\begin{figure}[htbp]
\centerline{\includegraphics[width=\columnwidth]{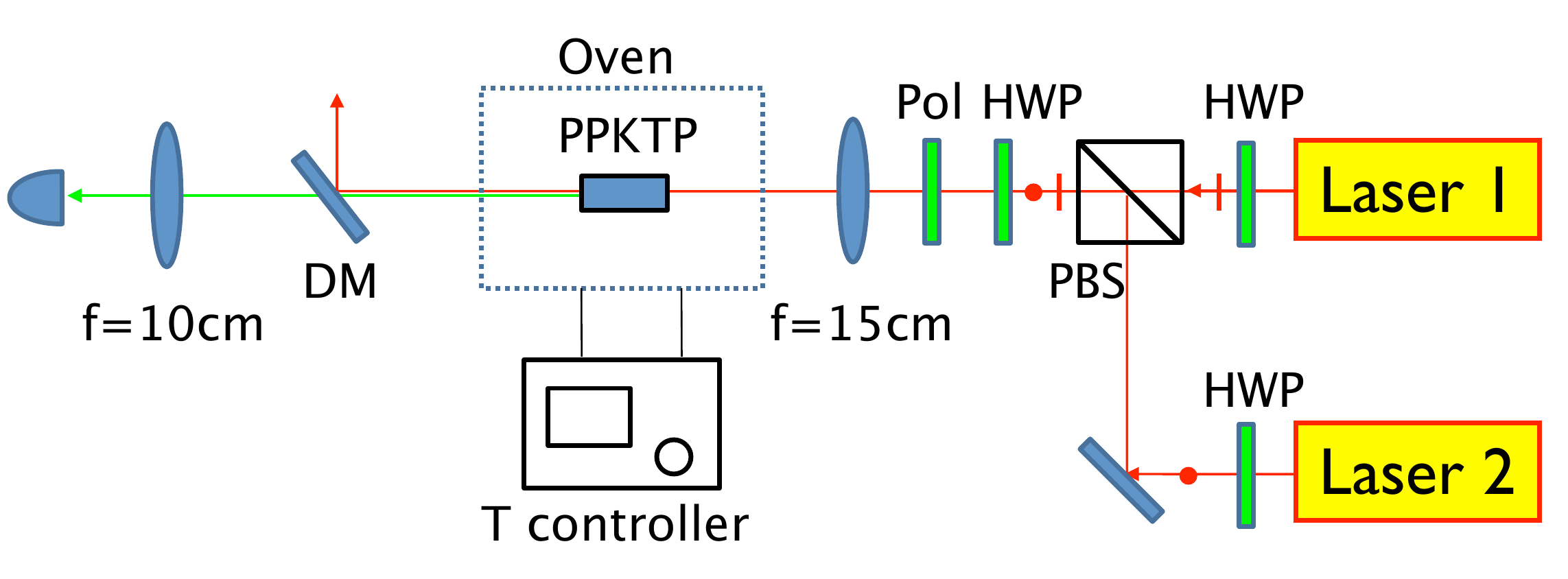}}
\caption{Experimental setup. HWP: half wave plate; PBS: polarizing beam splitter; Pol: polarizer; DM: dichroic mirror (HR for IR and AR for green).}
\label{fig:exp}
\end{figure}

In order to test this prediction, we used experimental SFG with 2 Newport-New Focus TLB-6721 ``Velocity'' tunable, continuous-wave, linearly polarized diode lasers, the sum of whose frequencies was kept constant over a 1058-1070 nm wavelength range. The experimental setup is shown in \fig{exp}. 
The x-cut, $1\times2\times10$ mm$^{3}$ PPKTP crystal was poled at a period of 9.0 $\mu$m (at room temperature) and provided by Raicol Crystals. The crystal faces were antireflection coated by Advanced Thin Films for both $y$ and $z$ polarizations at 1064 nm and for the $z$ polarization at 532 nm. The PPKTP crystal was  placed in a temperature-stabilized oven, controlled by a Peltier module driven by a Wavelength Electronics, Model LFI-3751 temperature controller.  

The power of the lasers are set to be 23.0(2) mW and 19.6(2) mW, respectively, to compensate the efficiency difference caused by their different beam profiles. The power of the $z$-polarized beam entering the crystal are 7.7(2) mW and 7.0(2) mW respectively. Both lasers were scanned in the wavelength range of 1058-1070 nm, keeping the average wavelength of the two lasers fixed at $1064.090(5)$ nm, so that the generated SFG green beam always had $\lambda= 532.05$ nm. The IR frequencies displayed by each laser were initially measured with a Coherent WaveMaster wavelength meter for calibration. 

\begin{figure*}[hbtp]
\centerline{\includegraphics[width=2\columnwidth]{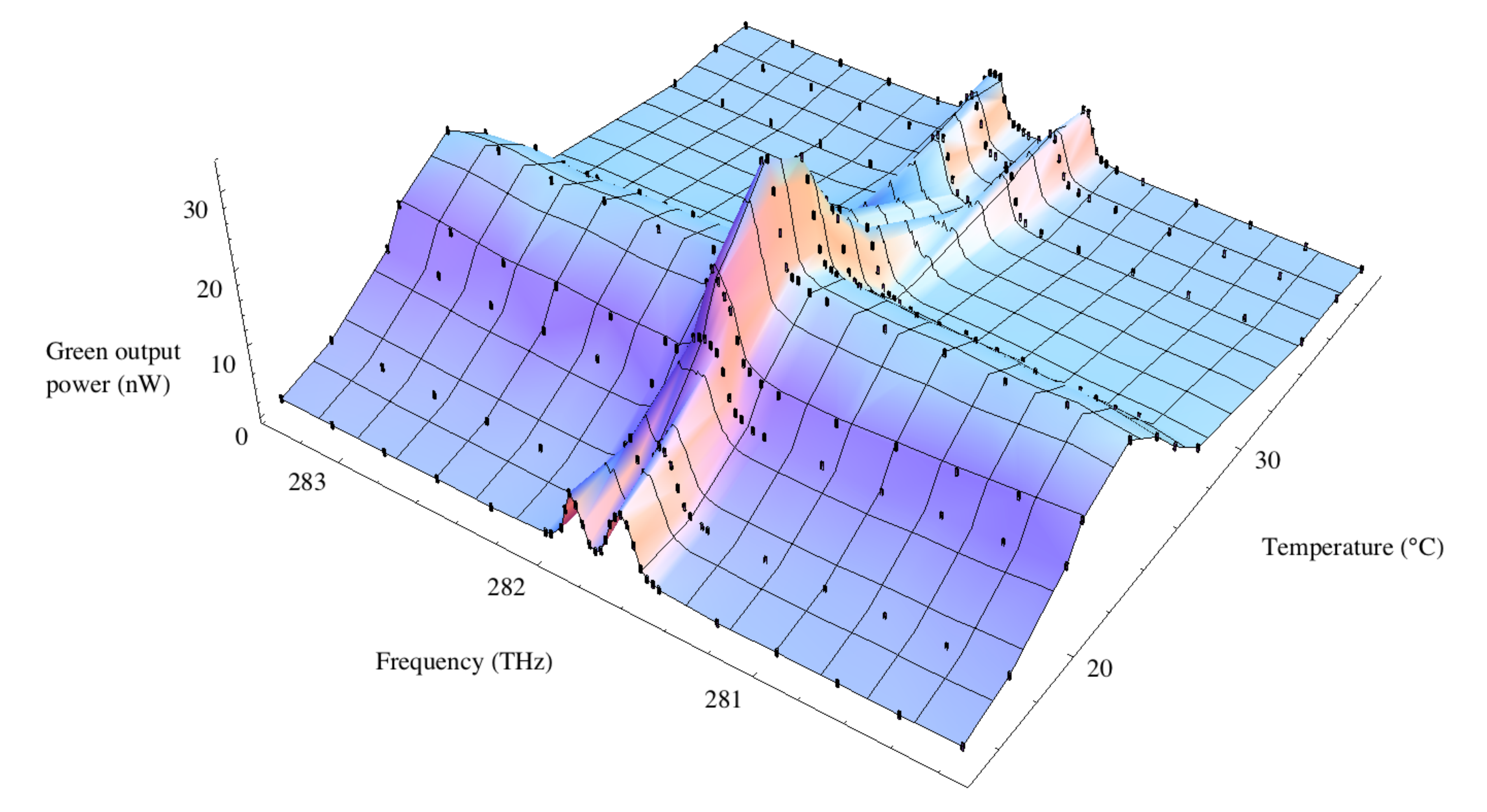}}
\centerline{\includegraphics[width=2\columnwidth]{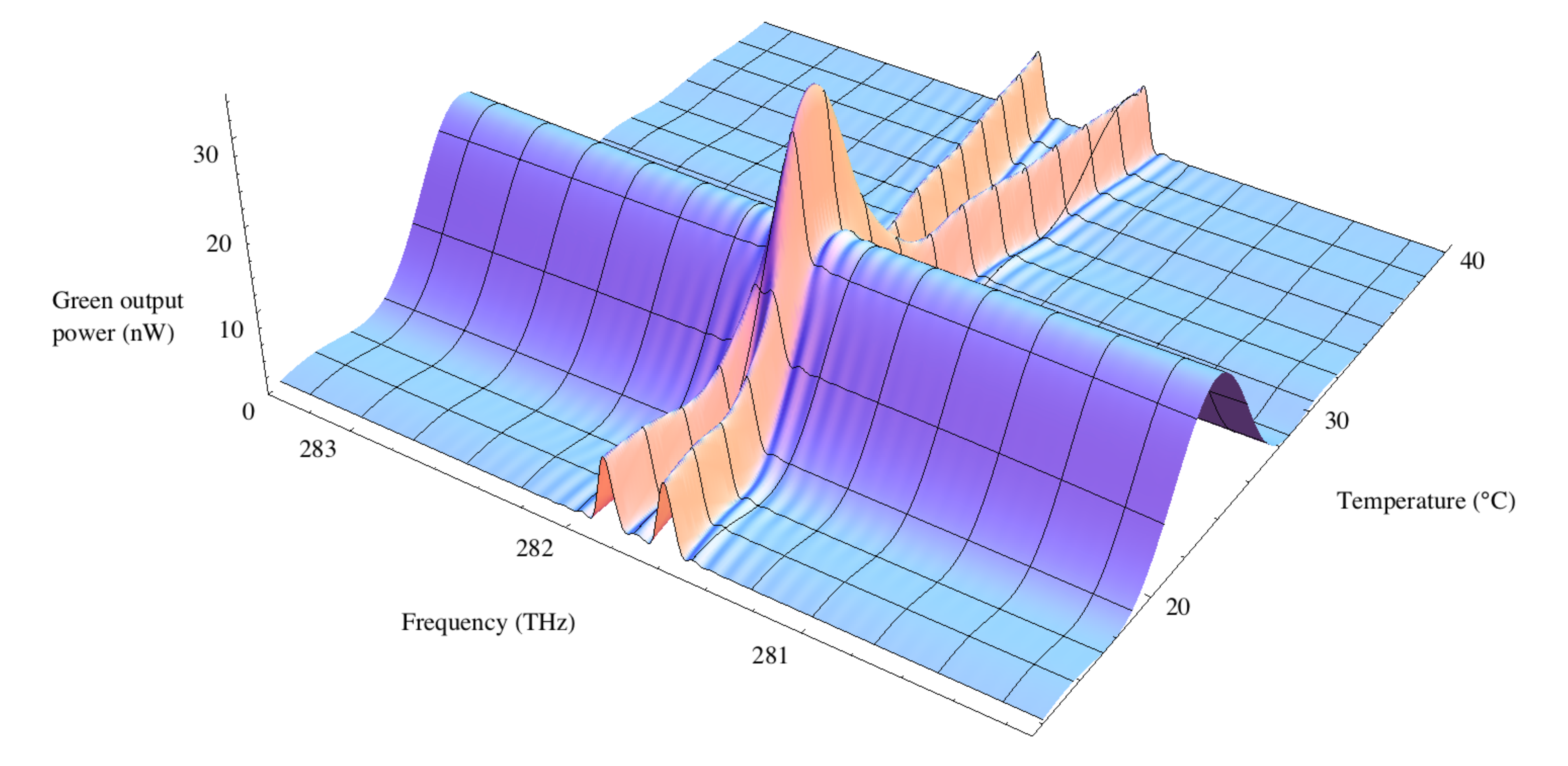}}
\caption{(a) Experimental phasematching curve. The laser wavelength is scanned from 1058 to 1070 nm, the temperature of the crystal is scanned from 15$^{\circ}$C to 40$^{\circ}$C (11 different temperatures). About 30 data points of different wavelengths were measured at each temperature. The 3D plot was obtained by a linear interpolation (Mathematica) from the data points. (b) Theoretical phasematching curve, plotted by using the respective power values   18.28 nW for $g_3^2L^2|a_1a_2|^2$, and 8.23 nW for  $g_1^2L^2|a_1|^4=g_2^2L^2|a_2|^4$ in \eq{int}.  
The measured SFG bandwidth is 3.178(2) THz, at quasi-constant efficiency, around 23$^{\circ}$C.  The crossing ridges are due to the more narrowly quasiphasematched SHG interactions.
}\label{fig:res}
\end{figure*}

\Fig{res}(a) shows the measured power of generated green beam (both SFG and SHG) versus the fundamental wavelength  and crystal temperature. For each temperature, about 30 data points of different wavelength are measured.
The maximum SFG and SHG power were $19.2(1)$ nW and $8.94(1)$ nW respectively, yielding maximum efficiencies for SFG and SHG of $3.60(2)\times 10^{-4}$ W$^{-1}$ and $1.67(1)\times 10^{-4}$ W$^{-1}$, respectively. We attributed these relatively modest values to the unoptimized overlap of the highly elliptical transverse intensity profiles of the diode lasers. However, we believe it is reasonable to assume that this imperfect overlap was stationary throughout the measurements.
 
\Fig{res}(b) shows the theoretical prediction for the total green intensity. As can be clearly seen, the qualitative behavior of SHG (frequency-sharp ridges) and SFG (broadband feature) match the experimental results very well. The measured frequency width of the constant SFG QPM was 3.178(2) THz, at least, since our measurement was limited by the tuning range of the diode lasers (we could not make use of the full 1030-1070 nm tuning range of the lasers as it was not centered on the SHG fundamental wavelength of 1064 nm). This measurement indicates that as many as 6700 cavity modes (beyond the 60 actually measured) could be entangled in our recent demonstration in a single OPO~\cite{Chen2014}. 

\section{OPO dispersion}

We now turn to the effect of the dispersion of the OPO nonlinear crystal. As we just showed in the previous section, for a pump at 532 nm, the QOFC can, in theory, span 4.74 THz, with at least 3.18 THz confirmed experimentally.  
However, due to the  dispersion of the nonlinear crystal of the OPO, we can expect the comb spacing to become uneven at frequencies far removed from half the pump frequency $\nu_o=\nu_p/2$. Let  $\Delta_{o}$ be the constant free spectral range (FSR) of the OPO, calculated at $\nu_{o}$ in the absence of dispersion. The emission of photon pairs by PDC into, and generation of EPR entanglement between, qumodes at the following frequencies placed symmetrically around $\nu_{o}$:
\begin{align}
\nu'_{\pm m}  &=\nu_o\pm\left( m-\frac12 \right)\Delta_{o}\\
&\equiv\nu_o+\mu_{\pm m}\Delta_{o},\label{eq:mum}
\end{align}
where $m$ is a  nonzero  integer. 

\begin{figure*}[htbp]
\centerline{\includegraphics[width=1.8\columnwidth]{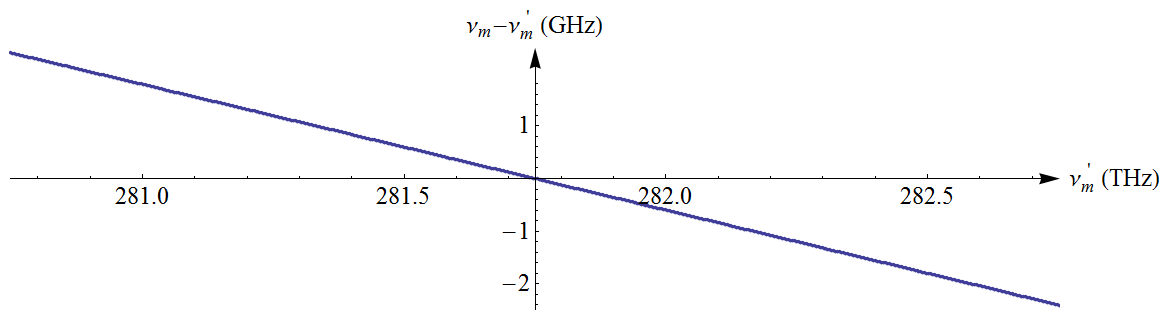}}
\caption{Frequency shift $\nu_m-\nu_m'$ vs $\nu_m'$, calculated by solving \eq{nuj} exactly. Within the plot range (2 THz) it is almost linear and symmetric around the center frequency $\nu_o$.}
\label{fig:fs}
\end{figure*}

In the presence of dispersion, the FSR will depend on frequency and will be denoted by $\Delta_{m}$ instead of $\Delta_{o}$, where $m$ is still taken to denote the $m^\text{th}$ mode from $\nu_{o}$. Let's consider the ring OPO used in Ref.~\citenum{Chen2014}, which contains two identical PPKTP crystals, oriented at 90$^\circ$ from each other. With this arrangement, the OPO cavity is polarization degenerate and the resonance condition for either linear principal polarization $y$ or $z$  of the PPKTP crystals is 
\begin{equation}\label{eq:rc}
\frac{2\pi\nu_{m}}c \left\{n_\text{air}(\nu_{m})L + [n_z(\nu_m)+n_y(\nu_m)]\ell\right\}  = j_{m} 2\pi,
\end{equation}
where $j_{m}$ is a positive integer, $c$ is the speed of light, $L$ is the length of the air path in a round trip, $\ell$ is the length of each PPKTP crystal, $n_\text{air}$ is the index of refraction of air, and $n_z$ and  $n_y$ are the refractive indices for polarizations along the $z$ and $y$ principal axes of KTP, respectively. The OPO resonance frequencies are 
\begin{equation}\label{eq:nuj}
\nu_m=  j_{m} \frac{c}{n_\text{air}(\nu_{m})L + [n_z(\nu_m)+n_y(\nu_m)]\ell}\equiv 
j_m\Delta_{m}.
\end{equation}
It is worth noting here that, because of the complex dependence of the left-hand side on $m$, 
it is not {\em a priori} warranted that two consecutive frequencies $\nu_{m}$ and $\nu_{m+1}$ should correspond to the respective values $j_{m}$ and $j_{m+1}$ such that 
\begin{equation}\label{eq:j}
j_{m+1}=j_{m}+1.
\end{equation}
However, as we will soon see (\fig{fs}),  the dispersion effects will always be monotonous enough that \eq j always applies. 

\begin{figure}[htbp]
\centerline{\includegraphics[width=\columnwidth]{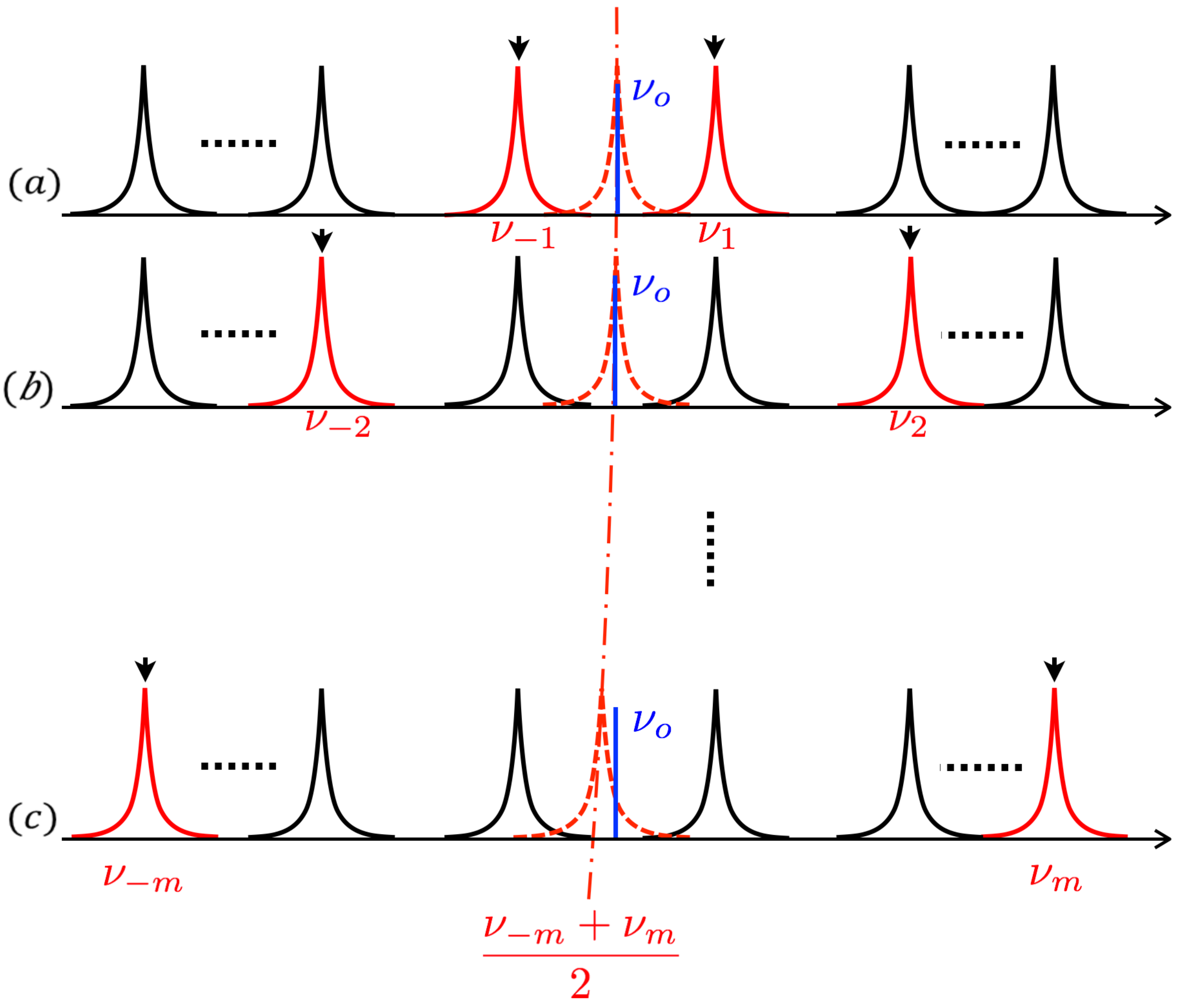}}
\caption{The dispersion effect of the OPO. The OPO pump has a fixed frequency and very narrow linewidth; the blue solid vertical line in the center marks half the pump frequency $\nu_{o}$. The red solid lines denote the mode pair indexed by $\pm |m|$ and the red dashed line their average. When dispersion is negligible ($m\sim 1$), the symmetric resonant modes $\pm m$ of the OPO are equally spaced by a quasi-constant FSR: (a), $m=1$, $\nu_{o}\simeq(\nu_{-1}+\nu_{1})/2$; (b), $m=2$, $\nu_{o}\simeq(\nu_{-2}+\nu_{2})/2$. When dispersion is nonnegligible ($m\gg1$), the resonant modes of the OPO (red solid lines) are not equally spaced by a constant FSR: (c), $\nu_{o}\neq(\nu_{-m}+\nu_{m})/2$, and the PDC modes do not overlap with OPO eigenmodes any more.}
\label{fig:avgshiftplot}
\end{figure}

The expected effect of normal dispersion is qualitatively depicted in \fig{avgshiftplot}: we expect the average frequency $(\nu_m+\nu_{-m})/2$ to veer away from $\nu_{o}$ as $m\gg1$, which will lead to a severe degradation of squeezing as the shift $(\nu_{m}+\nu_{-m})/2-\nu_{o}$ reaches the OPO linewidth. 

We now evaluate this effect quantitatively.  We can rewrite \eq{mum} as
\begin{equation}
\nu_m'= j_m\Delta_o,
\end{equation}
where 
\begin{equation}
j_{m} = \frac{\nu_{o}}{\Delta_{o}} +\mu_{m}\equiv\zeta_{o}+\mu_{m},
\end{equation}
which assumes \eq j holds. Note that $j_m$ {\em is} an integer because $\nu_{o}$ is always placed exactly between two consecutive OPO modes in our experiments~\cite{Chen2014}, which makes $\zeta_{o}$ a half integer, like $\mu_{m}$. 

We solved \eq{nuj} for each given $j_{m}$, using the Sellmeier equations of air~\cite{Fang2002} and KTP~\cite{Fradkin1999,Fan1987,Emanueli2003} 
and the parameters of  our OPO: $L = 28
$ cm, $l = 1 $ cm, $\Delta_o = 0.945$ GHz, $\lambda_{o} = 1064.04$ nm. 
We obtained the correspondent resonant frequency $\nu_m$ and the resulting difference  $\nu_m-\nu_m'$, the frequency shift due to dispersion, was plotted in \fig{fs},
which clearly shows that the frequency is indeed a monotonous function of $m$ in the presence of dispersion. This confirms that \eq j does apply for neighboring frequencies in the presence of dispersion.

\begin{figure*}[htbp]
\centerline{\includegraphics[width=1.8\columnwidth]{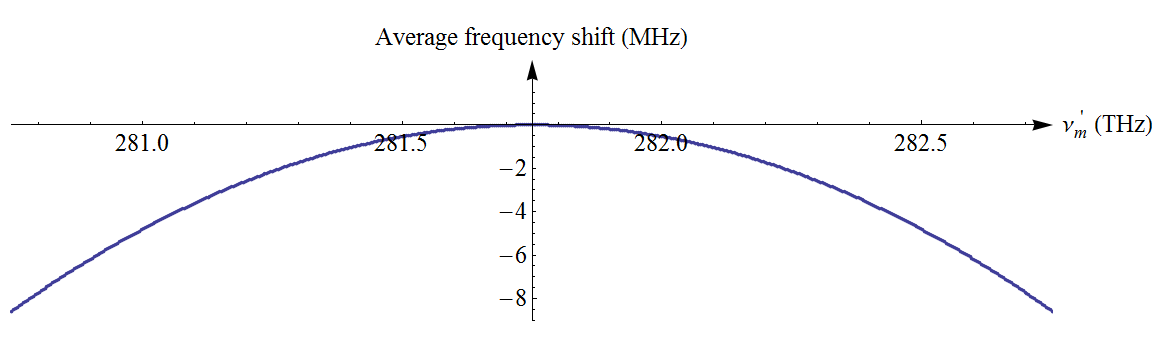}}
\caption{Average frequency shift vs $\nu_{j}'$, calculated by solving \eq{nuj} exactly.}
\label{fig:avFshiftEffect}
\end{figure*}

Of prime interest to us is the shift of the average frequency $(\nu_{m}+\nu_{-m})/2$ with respect to its dispersionless value $\nu_{o}$. If the plot of \fig{fs} is a straight line, this shift is exactly zero. However, the GHz scale of \fig{fs} doesn't allow us to discern whether a shift of the order of the crucial OPO linewidth (10 MHz) is actually present. We therefore plot the shift of the average explicitly in \fig{avFshiftEffect}.

As can be seen from \fig{avFshiftEffect}, the shift of the average becomes significant at about 1 THz away from $\nu_{o}$. This means that the PDC modes won't overlap any more with dispersion-chirped OPO resonances, as in the bottom row of \fig{avgshiftplot}, and the squeezing will drop even though we are still well within the quasiphasematching bandwidth.

A first-order Taylor expansion of \eq{nuj} provides a simple solution for the resonant frequencies with good precision. We only study the frequency range of 1  THz around $\nu_o$, where the dispersion of air is $10^{-5}$ of the dispersion of the crystal, so we  treat it as a constant: $n_{air}(\nu_m) = 1$. The Taylor expansion gives: 
\begin{align}
n_{z,y}(\nu_m) = n_{z,y}(\nu_o) &+ (\nu_m-\nu_o)\left.{dn_{z,y}\over d\nu}\right|_{\nu_o}  \nonumber\\
&+{(\nu_m-\nu_o)^2\over2}\left.{d^2n_{z,y}\over d\nu^2}\right|_{\nu_o} + \mathcal O(\nu_m^3).
\end{align}
It is straightforward to show that 
\begin{equation}
\left.{dn_{z,y}\over d\nu}\right|_{\nu_o} \approx \nu_o \left.{d^2n_{z,y}\over d\nu^2}\right|_{\nu_o},
\end{equation}
which yields
\begin{equation}
(\nu_m-\nu_o)\left.{d^2n_{z,y}\over d\nu^2}\right|_{\nu_o} 
\ll \left.{dn_{z,y}\over d\nu}\right|_{\nu_o}
\end{equation}
 when $\nu_m-\nu_o \ll \nu_o$. Therefore we can limit the expansion to the first order. The result is
\begin{equation}\label{eq:numd1}
\nu_m = j_m{c\over L + n_o \ell + \ell(\nu_m-\nu_o)\left.{d(n_z+n_y)\over d\nu}\right|_{\nu_o}}
\end{equation}
where $n_o = n_z(\nu_o) + n_y(\nu_o)$. Using the fact that
 \begin{equation}
\ell(\nu_m-\nu_o)\left.{d(n_z+n_y)\over d\nu}\right|_{\nu_o} \ll L + n_o \ell,
\end{equation}
we can simplify \eq{numd1} to:
\begin{align}
\nu_m &= j_m\Delta_o{1+{\ell\over c}\nu_o\Delta_o\left.{d(n_z+n_y)\over d\nu}\right|_{\nu_o}\over 1+ {\ell\over c}j_m\Delta_o^2\left.{d(n_z+n_y)\over d\nu}\right|_{\nu_o}}
\nonumber\\
&= j_m\Delta_o{1 - {\ell\over \zeta_o}\disp\over 1 - {j_m\ell\over \zeta_o^2}\disp}
\end{align}
where $\disp =\left.{d(n_z+n_y)\over d\lambda}\right|_{\lambda_0}$. 
Finally, we obtain
\begin{align}\label{eq:2order}
\nu_m 
 &= \nu_o + \mu_{m}\Delta' + \mu_{m}^2 {\Delta_o\ell\disp\over \zeta_o^2}
\end{align}
where $\Delta' = \Delta_o(1 + {\ell\over \zeta_o}\disp)$. This equation gives a  good approximate solution of \eq{nuj}, whose exact solution plotted on \figs{fs}{avFshiftEffect}.

The linear term in $\mu_{m}$, i.e., $\nu'_{m}$ is clearly dominant, as can be seen from the plot of the exact solution in \fig{fs}. This term shows that the resonant modes are still equally spaced, but the spacing is $\Delta'$, instead of $\Delta_o$ in the nondispersive case. For normal dispersion, $\disp < 0$, so  $\Delta' < \Delta_o$, which means that  the resonant modes move toward  the center frequency $\nu_o$, however all average frequencies are unshifted: $(\nu_m+\nu_{-m})/2=\nu_{o}$. Note that this conclusion does not depend on the choice of $\nu_o$. 
 
The second-order term in \eq{2order} is much smaller and yields the shift of the average frequency away from $\nu_{o}$. As mentioned earlier, this shift is significant because it can reach the order of the OPO qumode linewidth. \Eq{2order} gives
\begin{equation}\label{eq:approxAvg}
\frac{\nu_{-m}+\nu_{m}}2 - \nu_o =  \mu_{m}^2 {\Delta_o\ell\disp\over \zeta_o^2}.
\end{equation}
Typically, $ \zeta_o = 3\times10^5$, $\disp = - 7\times10^4$ m$^{-1}$, which gives
\begin{equation}\label{eq:m}
\frac{\nu_{-m}+\nu_{m}}2 - \nu_o = -7.77\ m^2,
\end{equation}
in hertz. We can invert \eq m to obtain the maximum number of entangled qumodes $m$ for which the shift of the average equals the OPO half width at half maximum of 5 MHz (the 1 kHz pump linewidth is negligible). This is obtained for $m \sim 800$. At this value, the deviation from the exact solution is only 25 kHz. Our theoretical prediction of the number of entangled qumodes of both polarizations, as limited by dispersion in the experiment of Ref.~\citenum{Chen2014}, is therefore 3200 qumodes.

It is important to note that dispersion compensation can alleviate this effect and restore the quasiphasematching bandwidth of (at least) 6700 entangled qumodes. It should be noted, This dispersion management has been mastered for femtosecond frequency combs, using prism compressors~\cite{gale1995,wachman1990}, chirped mirrors~\cite{matuschek1999,hebling1995},  opposite group velocity dispersion materials~\cite{leindecker2011,leindecker2012}. However, it might be more challenging in our case because it involves quite a narrower frequency span.

\section{Conclusion}
In conclusion, we have investigated the experimental limits of the record-size, 60-qumode cluster state of our previous work in Ref.~\citenum{Chen2014}. Whereas the 60-qumode figure arose from a technical limitation in the measurement of the entanglement (limited tunability of the local oscillator), we expected that the experimental limitations in the actual generation of entanglement to be much laxer. Here, we confirmed these  expectations. We first demonstrated broadband single-period quasiphasematching of sum-frequency generation, i.e., also of parametric downconversion, in $9\mu$m-poled PPKTP for 532/1064 nm up/down-converted wavelengths, then studied the OPO dispersion's effect on the PDC. The 3.178(2) THz measured QPM bandwidth, at constant SFG efficiency, was limited by the tunability of our diode lasers and hence this result is only a lower bound for the actual QPM bandwidth. (Note that even broader phasematching has been achieved for SFG, using noncollinear beams in birefringently phasematched LBO~\cite{Peer2007}, and for parametric dowconversion, using chirped quasiphasematching with collinear~\cite{Harris2007} and noncollinear~\cite{Tanaka2012} propagation.) We discovered a flat-top quasiphasematching feature in a simple single-period PPKTP crystal, limiting ourselves to the characterization of the experimental situation of Refs.~\citenum{Pysher2011} and~\citenum{Chen2014}, which requires collinear modes, all uniformly coupled and matched to the same OPO cavity. As it stands, this measurement indicates that as many as 6700 cavity modes (beyond the 60 actually measured) could be entangled in our recent demonstration in a single OPO~\cite{Chen2014}, if dispersion in the KTP played no role. 

When dispersion is present, we showed that the QOFC gradually becomes more unevenly spaced away from the center frequency (half of the pump frequency), which leads to decreasing overlap between the PDC modes and the OPO resonant modes and loss the resonant enhancement of qumode entanglement by the OPO cavity. We calculated that the limit is 3200 entangled qumodes in Ref.~\citenum{Chen2014}. Successful dispersion management will bring back the quasiphasematching limit of --- at least ---  6700 qumodes.

\section*{Acknowledgments}
We thank Moran Chen for discussions. This work was supported by the U.S. National Science Foundation under grant No.\ PHY-1206029, by the University of Virginia Distinguished Research Award, and by the Higher Education Equipment Trust Fund of the State of Virginia. 

%

\end{document}